\documentclass[runningheads]{llncs}
\usepackage{graphicx}
\usepackage{cite}
\begin{document}
\title{Incorporating Improved Sinusoidal Threshold-based Semi-supervised Method and Diffusion Models for Osteoporosis Diagnosis}
%
%
\author{Ke Wenchi}
%
%
\institute{sichuan university\\
\email{kewenchi26@gmail.com}\\}
\maketitle              
\begin{abstract}
Osteoporosis is a common skeletal disease that seriously affects patients' quality of life. Traditional osteoporosis diagnosis methods are expensive and complex. The semi-supervised model based on diffusion model and class threshold sinusoidal decay proposed in this paper can automatically diagnose osteoporosis based on patient's imaging data, which has the advantages of convenience, accuracy, and low cost. Unlike previous semi-supervised models, all the unlabeled data used in this paper are generated by the diffusion model. Compared with real unlabeled data, synthetic data generated by the diffusion model show better performance. In addition, this paper proposes a novel pseudo-label threshold adjustment mechanism, Sinusoidal Threshold Decay, which can make the semi-supervised model converge more quickly and improve its performance. Specifically, the method is tested on a dataset including 749 dental panoramic images, and its achieved leading detect performance and produces a 80.10\% accuracy. 
\end{abstract}
\section{Introduction}
\par Osteoporosis is a chronic medical condition that affects the bones, leading to a loss of bone density and an increased risk of fractures. It is a common condition, affecting one in three women and one in five men over the age of 50 worldwide\cite{1,2,3}, particularly among post-menopausal women and older adults, and can have a significant impact on quality of life. According to the International Osteoporosis Foundation, osteoporosis is estimated to affect 200 million women - approximately one-tenth of women aged 60, one-fifth of women aged 70, two-fifths of women aged 80 and two-thirds of women aged 90. In Europe, the disability due to osteoporosis is greater than that caused by cancers (with the exception of lung cancer) and is comparable or greater than that lost to a variety of chronic noncommunicable diseases, such as rheumatoid arthritis, asthma and high blood pressure related heart disease\cite{4}. Due to its prevalence worldwide, osteoporosis is considered a serious public health concern\cite{5}. 
\par However, bone density testing can be expensive and may not always be covered by insurance. Some bone density tests require patients to lie still for extended periods, which can be difficult for some people. In certain areas, access to bone density testing may be limited, making it difficult for patients to undergo regular screening. using computer-aided diagnosis, doctors can diagnose osteoporosis more quickly and accurately, helping patients receive timely treatment, reducing the risk of fractures, and improving their quality of life.
\par Dual-energy X-ray absorptiometry (DXA) is an effective means of identifying bone mineral density (BMD) and is the standard test for diagnosing osteoporosis. However, DXA scans are relatively expensive, making them unsuitable for general screening. Dental panoramic radiographs are cheaper. In past literature, several indicators, including mandibular cortical width (MCW) and mandibular cortical index (MCI)\cite{6}, have been used for diagnosing osteoporosis using dental panoramic radiographs. However, these diagnostic methods are not widely used in clinical settings as they are too complex for manual interpretation and require a high level of skill.
\subsection{Relate work}
Previous methods for diagnosing osteoporosis mainly relied on manually classified feature indicators. In recent years, deep learning techniques have made great progress in computer-aided diagnosis and have been applied to the diagnosis of osteoporosis. For example, Tang\cite{tang2021cnn} used a convolutional neural network-based technique to screen for osteoporosis on 2D CT slices of the lumbar spine, while Wani\cite{wani2022osteoporosis} designed a convolutional neural network-based osteoporosis diagnosis scheme that can screen for knee joint X-rays. Some research groups have proposed solutions for panoramic dental images, such as Lee's\cite{7} transfer learning-based method for osteoporosis screening and Adillion's\cite{8} use of Line Operator (LO) as a preprocessing method to enhance the performance of CNN-based osteoporosis detection. These studies have achieved certain results using convolutional neural network models on their respective datasets. However, the dataset sizes of these studies are quite small, which may lead to generalization issues. It is worth noting that clinical data collection is often difficult, so the performance of supervised models may not reach their optimal performance.
\par In this paper, we propose a novel semi-supervised training framework for osteoporosis diagnosis using complete oral panoramic radiographs as input. Specifically, we introduce the latest generative diffusion models to generate a large amount of synthetic data from real unlabeled data, which is then fed into the model. Using the latest generative diffusion models, we generate a large amount of synthetic data and use it to train our semi-supervised model. Surprisingly, compared to using real unlabeled data, using synthetic data can achieve better results. To further improve experimental outcomes, we analyze the problem of threshold setting in existing methods and propose a new semi-supervised confidence threshold adjustment method, which helps improve convergence speed and increase the final accuracy of osteoporosis diagnosis.

\section{Method}
\subsection{Materials}
Our dataset was collected from * Hospital, with age ranging from 16 to 90 years old. This retrospective study was approved by the Ethics Committee of * University. The OPGs were obtained from patients who underwent routine clinical evaluation between 20* and 201*. Our data was selected based on age, and only individuals above 45 years old were included in our study sample. After being labeled by experienced oral physicians, these samples were divided into three categories according to the mandibular cortical index (MCI), which were C1 with basic lesions, C2 with moderate lesions, and C3 with severe lesions\cite{6}. The training set is 399 labeled images, 133 images for each category and 12000 unlabeled images. The test set has a total of 749 images, about 250 images in each category.
\par In previous literature, image preprocessing for medical diagnosis can be complicated and may pose some limitations on algorithm deployment. To minimize the problems caused by preprocessing, we did not perform any special preprocessing on the images. We only converted them to a usable lossless image format and then resized the images to fit the network input. Other than that, no other data processing was performed.
\subsection{Architecture}
In this study, a standard semi-supervised training process was used, as shown in Figure 1. After simple preprocessing, the unlabeled data in the dataset enters a generative model to generate a large amount of unlabeled simulation data. The generated simulation data and real labeled data are then sent to semi-supervised model for training. The backbone we used is a wideresnet network\cite{zagoruyko2016wide}. In order to improve the results of semi-supervised training, we designed a threshold adjustment module based on sine decay, referring to the idea of pseudo-label learning in semi-supervised learning. The sine decay threshold adjustment module can dynamically adjust the threshold throughout the entire training phase, so that the consistency loss is constrained to a reasonable range, resulting in significantly improved convergence speed and classification results of the model training.
\begin{figure} 
\centering
\includegraphics[width=0.9\textwidth]{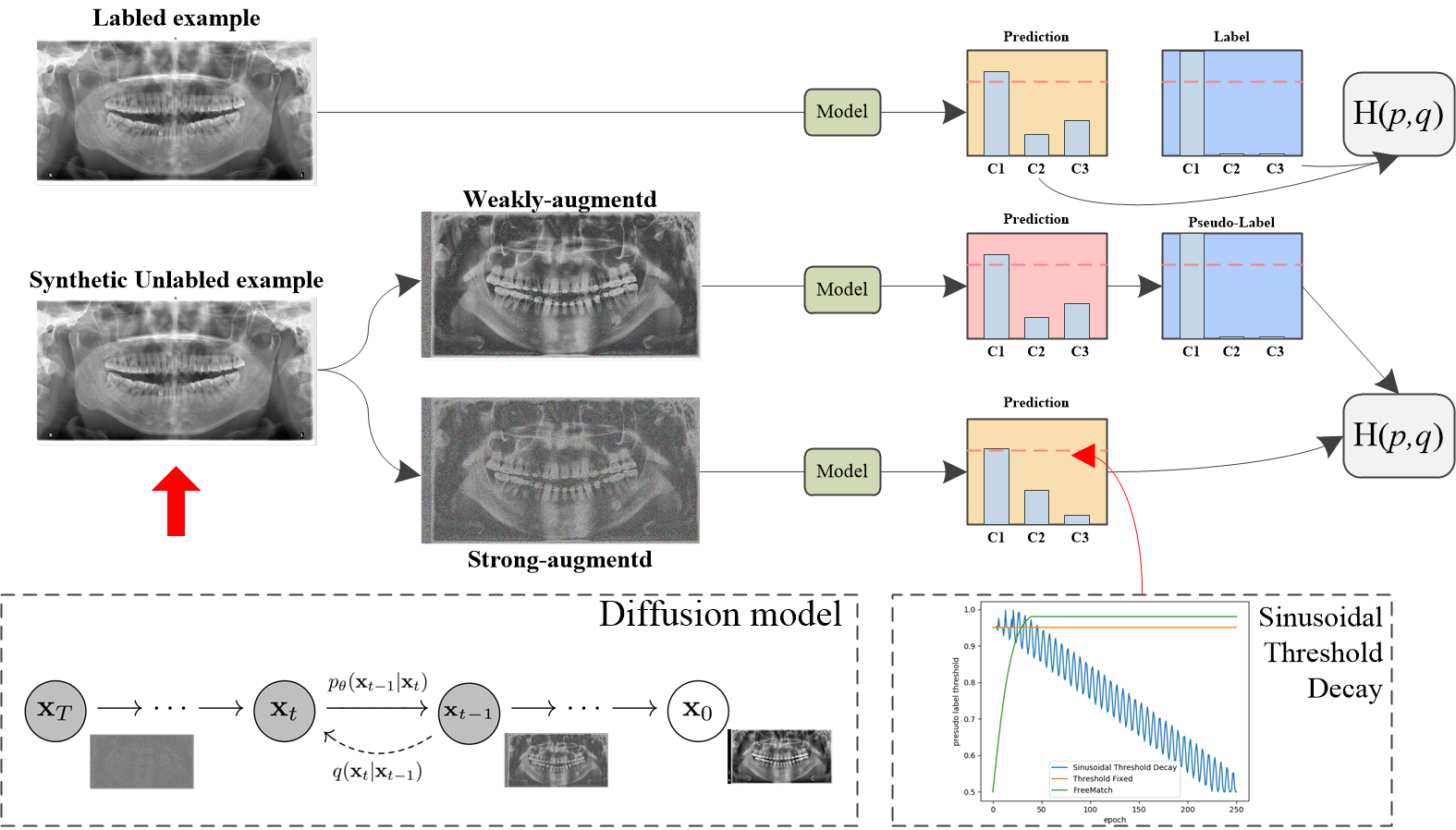} \caption{The overall Architecture of the method in this paper.} \label{fig1} \end{figure}

\subsection{Generating Synthetic Data Based on Denoising Diffusion Probabilistic Models(DDPM)}
\par Diffusion Models\cite{ho2020denoising} is a type of generative model that generates high-quality synthetic data. The main idea of diffusion models is to take a real data point and generate a distribution by repeatedly adding random noise, and then sample from the distribution to generate synthetic data points. Using diffusion models can generate highly realistic data because the perturbation process gradually simulates the complex structure of real data at each step. In recent research, diffusion-based methods have achieved excellent performance in many fields, such as image generation, speech generation, natural language processing, etc.
\par Naturally, we apply diffusion models to the generation of synthetic panoramic dental X-ray images. In this study, we used the model in \cite{ho2020denoising,goodfellow2014generative} to generate synthetic data using 12,000 unlabeled real panoramic dental X-ray images. The model's generation parameters can be found in the appendix. These generated synthetic images will be considered as the next stage of unlabeled data and will be used for semi-supervised training. It should be noted that the real data used to generate the synthetic data does not contain labels, and is not included in the labeled dataset and the test set, so there is no potential data leakage. As shown in Figure 2, we were surprised by the outstanding performance of the diffusion model in generating panoramic dental X-ray images, which is almost indistinguishable from real data. To better evaluate the effectiveness of the diffusion model, we invited several professional dentists to evaluate the authenticity of the generated synthetic data. The result was a staggering the synthetic data generated by the diffusion model is so realistic that even dental professionals can hardly tell whether it is real or fake.

\begin{figure} 
\centering
\includegraphics[width=0.9\textwidth]{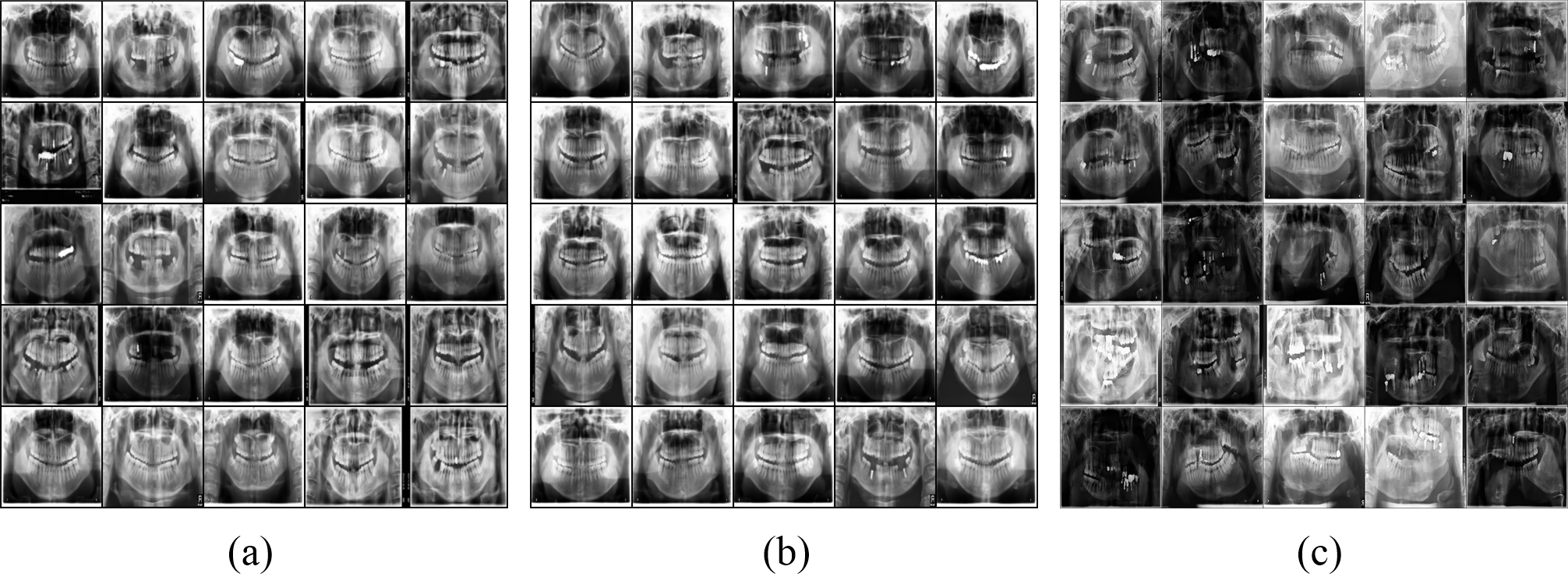} \caption{Results of synthetic data generated by different models. From left to right: a) Denoising Diffusion Probabilistic Model(DDPM); b) real data ;c)Generative Adversarial model(GAN).} \label{fig2} \end{figure}

\subsection{Semi-supervised methods based on simulated data.}
\par In computer-aided diagnosis systems, insufficient available data is often encountered. The reason for this problem is that there is a large amount of disease data, but it is difficult to label them all. Hence, using semi-supervised learning\cite{sohn2020fixmatch} is an effective method to improve the efficiency of disease diagnosis. In this paper, We trained models using both real and simulated unlabeled data, and the results were very interesting, as shown in Table 2. We conducted experiments with different amounts of unlabeled data, and it can be seen that the final model performance using real unlabeled data is actually worse than using simulated data, and it is significantly behind. Furthermore, we found in the table that using real unlabeled data, as the amount of data increases, actually inhibits the final model performance. Using semi-supervised strategies for medical image diagnosis can lead to a decrease in performance. This phenomenon is widely observed in many unpublished works.
\par Based on the experimental results in this paper, a possible reasonable explanation is that there may exist some data bias in the real data. If we view an image as a specific feature vector in a high-dimensional space, the distribution of real data is not uniform. Sampling in the real world often struggles to overcome this non-uniformity. Using generated simulation data can be seen to some extent as interpolating between discrete feature vectors in a high-dimensional space, making the data distribution more uniform. However, this premise is based on the foundation that our simulation data is realistic enough. If the generated model is not good enough to produce sufficiently complex details and textures, it can actually reduce the performance of the algorithm.
\subsection{Sinusoidal Threshold Decay for Consistency Learning}
\par This paper proposes an improved method for the threshold of pseudo-labels in consistency-based semi-supervised learning\cite{sohn2020fixmatch}, which not only enhances the convergence speed but also improves the final model performance. In semi-supervised learning, there is a crucial parameter called the pseudo-label threshold. This threshold significantly affects the final performance of the model and controls the consistency loss. FreeMatch\cite{wang2022freematch} proposed an adaptive threshold adjustment method, gradually increasing the threshold from a low value to a high level. In contrast, this paper presents a completely opposite strategy, gradually decreasing the threshold from a high level to a reasonable level.
\par Inspired by the pseudo-label-based semi-supervised method\cite{rizvedefense,liu2022acpl}, we argued that a fixed threshold is unreasonable, and gradually increasing the threshold from a low value is also unreasonable. The pseudo-label threshold plays a crucial role in semi-supervised learning by determining whether two samples belong to the same class. If a high threshold is not set at the beginning of training, the model may misclassify many low-confidence samples, potentially leading to non-convergence problems. Maintaining a high threshold throughout training prevents many samples from being excluded from training, resulting in severe overfitting problems with increasing training time.
\begin{figure} 
\centering
\includegraphics[width=0.65\textwidth]{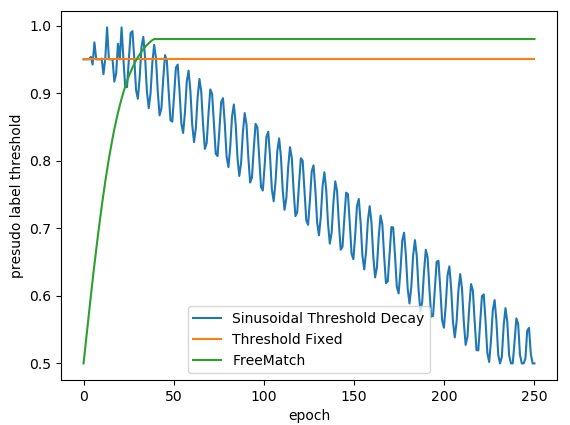} \caption{Visualized threshold changes.} \label{fig3} 
\end{figure}
\par Therefore, based on the above reasons, this paper proposes a sine threshold descent algorithm. The threshold gradually decreases with the number of iterations, with only a few strong samples correctly identified as belonging to the same class at the beginning. As the number of iterations increases, the model's performance improves, and many weak and strong samples are included in the training process. A smooth descent threshold is not a good idea. The threshold value for the same sample pair during training is not a fixed value but corresponds to a certain range. Therefore, in this paper, a sine wave will be added to superimpose fluctuations of the threshold within a finite range during the descent process, which helps the model obtain stronger robustness. The threshold change strategy is shown in Figure 2.
\par The original threshold for pseudo-labeling $T_{f}$, where $i$ is the current iteration number and $i_{max}$ is the maximum iteration number. $\alpha$ is the minimum threshold constraint, set to 0.5 in this paper, and $\beta$ is the fluctuation coefficient that determines the degree of fluctuation during the descent process, set to 0.05 in this paper. The sine threshold $T_{s}$ can be defined as:
\begin{equation}
\mathrm{T}_{\mathrm{s}}=\mathrm{T}_{\mathrm{f}} \cdot \frac{\left(\mathrm{i}_{\max }-\mathrm{i} \cdot \alpha\right)}{\mathrm{i}_{\max }}+\beta \cdot \sin (\mathrm{i})
\end{equation}
\section{Experiments and Results}
\subsection{Evaluation Metrics and Implementation.}
\par Essentially, the diagnosis of osteoporosis is a classification problem, and we use accuracy as the evaluation metric. The framework is implemented based on PyTorch and utilizes an Nvidia 2080ti GPU. SGD optimizer is used, and the batch size is set to 8. The implementation will be available at ***.
\subsection{compared with other methods}
\par To evaluate the performance of our algorithm, we compared it with other state-of-the-art methods, as shown in Table 1. Since there is no publicly available panoramic dental radiograph dataset for osteoporosis, a fair comparison cannot be made. As the works in the table do not disclose their implementation details, we did not reproduce their methods in this paper but directly used the reported results from their papers. Additionally, as there are not enough osteoporosis diagnostic methods based on panoramic radiographs, we also reproduced common network models on the dataset used in this paper.
\begin{table}
\centering
\caption{Results compared to other methods}\label{tab1}
\begin{tabular}{|c|c|c|}
\hline
Method             & Accuracy(\%) & Testset Size \\ \hline
Adillion\cite{8}           & 87.5         & 24           \\
Lee\cite{7}                & 84.0         & 136          \\
Tassoker\cite{tassoker2022comparison}           & 81.14        & 148          \\
Vgg16\cite{simonyan2014very}              & 71.29        & 749          \\
Resnet(50)\cite{he2016deep}         & 74.10        & 749          \\
Densnet\cite{huang2017densely}            & 67.42        & 749          \\
wideresnet\cite{zagoruyko2016wide}         & 75.30        & 749          \\
\textbf{Proposed*} & 80.06        & \textbf{749} \\ \hline
\end{tabular}
\end{table}

\subsection{Simulation data or real data}
\par Before the semi-supervised model, we trained a generative diffusion model using unlabeled real data and generated 45,000 panoramic dental X-ray images. In the semi-supervised training phase, we conducted two different experiments, one using only real data and the other using only generated data, and compared them at different levels of labeled and unlabeled data. The experimental results are shown in Table 2. It can be seen from the table that when the number of labeled data used in training is small, both real and generated data can improve the accuracy of the model. However, comparing real data with generated data, real data is far inferior to generated data in terms of performance on any scale of labeled dataset. In addition, there is a surprising phenomenon that increasing real data will actually lower the performance of the model when there is a lot of labeled data.
\begin{table}[]
\caption{Results using simulated and real data}
\centering
\begin{tabular}{|c|cccc|}
\hline
Experiment             & \multicolumn{4}{c|}{Accurcy in different labeled data size}                                  \\ \cline{2-5} 
                       & \multicolumn{1}{c|}{399}   & \multicolumn{1}{c|}{300}   & \multicolumn{1}{c|}{201}   & 48    \\ \hline
Simulation data(100\%) & \multicolumn{1}{c|}{78.48} & \multicolumn{1}{c|}{76.58\%} & \multicolumn{1}{c|}{74.68\%} & 75.32\% \\
Simulation data(10\%)  & \multicolumn{1}{c|}{77.85} & \multicolumn{1}{c|}{75.43\%} & \multicolumn{1}{c|}{70.10\%} & 56.95\% \\
Real data(100\%)       & \multicolumn{1}{c|}{60.13} & \multicolumn{1}{c|}{59.27\%} & \multicolumn{1}{c|}{55.40\%} & 52.53\% \\
Real data(10\%)        & \multicolumn{1}{c|}{67.42} & \multicolumn{1}{c|}{64.48\%} & \multicolumn{1}{c|}{66.48\%} & 39.56\% \\ \hline
\end{tabular}
\end{table}
\subsection{Effectiveness of Sinusoidal Threshold Decay}
\par To further improve the diagnosis of osteoporosis, we proposed a threshold adjustment strategy based on a sine attenuation function in the paper. In this section, we verified the effectiveness of this strategy. It should be noted that all the unlabeled data used in the experiments were generated by a diffusion model, and the size of the labeled data was 399, with an average of 133 images per category. We discussed the cases of using only threshold descent, introducing sine wave fluctuation, and compared them with FreeMatch. The results are shown in Table 3. From the table, it can be observed that using FreeMatch, a threshold ascending method, did not achieve significant improvement in our task, and using only threshold descent did not achieve significant improvement either. The greatest improvement came from the introduction of sine wave fluctuation.
\begin{table}[]
\caption{Effectiveness of Sinusoidal Threshold Decay}
\centering
\begin{tabular}{|c|c|}
\hline
Experiment                                   & Accurcy(\%)    \\ \hline
Fix - Simulation data(100\%)\cite{sohn2020fixmatch}           & 78.48          \\
FreeMatch threshold ascent\cite{wang2022freematch}                    & 78.37          \\
Threshold Decay without Sinusoidal fluctuate & 78.77          \\
\textbf{Sinusoidal Threshold Decay*}          & \textbf{80.10} \\ \hline
\end{tabular}
\end{table}

\section{Conclusion}
\par In this work, we propose an improved semi-supervised approach using a sinusoidal threshold and a diffusion model for osteoporosis diagnosis. This method utilizes unlabeled data to generate a large amount of simulated data and completes the osteoporosis diagnosis under the semi-supervised training strategy. Our approach achieves superior performance and shows promise in our task. Additionally, our experiments demonstrate that sufficiently realistic simulated data can achieve better results than using real data directly. The introduced sinusoidal threshold decay method can improve the convergence speed of the model and enhance the final performance.

%
%
%
\bibliographystyle{splncs04}
\bibliography{ref}

\begin{thebibliography}{10}
\providecommand{\url}[1]{\texttt{#1}}
\providecommand{\urlprefix}{URL }
\providecommand{\doi}[1]{https://doi.org/#1}

\bibitem{8}
Adillion, I.G., Ishida, Y., Arifin, A.Z.: Line operator as preprocessing method
  for cnn-based osteoporosis detection in dental panoramic radiograph. In:
  Proceedings of the 6th International Conference on Frontiers of Educational
  Technologies. pp. 103--107 (2020)

\bibitem{goodfellow2014generative}
Goodfellow, I., Pouget-Abadie, J., Mirza, M., Xu, B., Warde-Farley, D., Ozair,
  S., Courville, A., Bengio, Y.: Generative adversarial nets in advances in
  neural information processing systems (nips). Curran Associates, Inc. Red
  Hook, NY, USA pp. 2672--2680 (2014)

\bibitem{he2016deep}
He, K., Zhang, X., Ren, S., Sun, J.: Deep residual learning for image
  recognition. In: Proceedings of the IEEE conference on computer vision and
  pattern recognition. pp. 770--778 (2016)

\bibitem{ho2020denoising}
Ho, J., Jain, A., Abbeel, P.: Denoising diffusion probabilistic models.
  Advances in Neural Information Processing Systems  \textbf{33},  6840--6851
  (2020)

\bibitem{huang2017densely}
Huang, G., Liu, Z., Van Der~Maaten, L., Weinberger, K.Q.: Densely connected
  convolutional networks. In: Proceedings of the IEEE conference on computer
  vision and pattern recognition. pp. 4700--4708 (2017)

\bibitem{4}
Johnell, O., Kanis, J.: An estimate of the worldwide prevalence and disability
  associated with osteoporotic fractures. Osteoporosis international
  \textbf{17},  1726--1733 (2006)

\bibitem{1}
Kanis, J., Johnell, O., Oden, A., Sernbo, I., Redlund-Johnell, I., Dawson, A.,
  De~Laet, C., Jonsson, B.: Long-term risk of osteoporotic fracture in
  malm{\"o}. Osteoporosis international  \textbf{11},  669--674 (2000)

\bibitem{6}
Klemetti, E., Kolmakov, S., Kr{\"o}ger, H.: Pantomography in assessment of the
  osteoporosis risk group. European Journal of Oral Sciences  \textbf{102}(1),
  68--72 (1994)

\bibitem{7}
Lee, K.S., Jung, S.K., Ryu, J.J., Shin, S.W., Choi, J.: Evaluation of transfer
  learning with deep convolutional neural networks for screening osteoporosis
  in dental panoramic radiographs. Journal of clinical medicine  \textbf{9}(2),
  ~392 (2020)

\bibitem{liu2022acpl}
Liu, F., Tian, Y., Chen, Y., Liu, Y., Belagiannis, V., Carneiro, G.: Acpl:
  Anti-curriculum pseudo-labelling for semi-supervised medical image
  classification. In: Proceedings of the IEEE/CVF Conference on Computer Vision
  and Pattern Recognition. pp. 20697--20706 (2022)

\bibitem{2}
Melton~III, L.J., Atkinson, E.J., O'Connor, M.K., O'Fallon, W.M., Riggs, B.L.:
  Bone density and fracture risk in men. Journal of Bone and Mineral Research
  \textbf{13}(12),  1915--1923 (1998)

\bibitem{3}
Melton~III, L.J., Chrischilles, E.A., Cooper, C., Lane, A.W., Riggs, B.L.:
  Perspective how many women have osteoporosis? Journal of bone and mineral
  research  \textbf{7}(9),  1005--1010 (1992)

\bibitem{5}
Peck, W.: Consensus development conference: diagnosis, prophylaxis, and
  treatment of osteoporosis. Am J Med  \textbf{94}(6),  646--650 (1993)

\bibitem{rizvedefense}
Rizve, M.N., Duarte, K., Rawat, Y.S., Shah, M.: In defense of pseudo-labeling:
  An uncertainty-aware pseudo-label selection framework for semi-supervised
  learning. In: International Conference on Learning Representations

\bibitem{simonyan2014very}
Simonyan, K., Zisserman, A.: Very deep convolutional networks for large-scale
  image recognition. arXiv preprint arXiv:1409.1556  (2014)

\bibitem{sohn2020fixmatch}
Sohn, K., Berthelot, D., Carlini, N., Zhang, Z., Zhang, H., Raffel, C.A.,
  Cubuk, E.D., Kurakin, A., Li, C.L.: Fixmatch: Simplifying semi-supervised
  learning with consistency and confidence. Advances in neural information
  processing systems  \textbf{33},  596--608 (2020)

\bibitem{tang2021cnn}
Tang, C., Zhang, W., Li, H., Li, L., Li, Z., Cai, A., Wang, L., Shi, D., Yan,
  B.: Cnn-based qualitative detection of bone mineral density via diagnostic ct
  slices for osteoporosis screening. Osteoporosis International  \textbf{32},
  971--979 (2021)

\bibitem{tassoker2022comparison}
Tassoker, M., {\"O}zi{\c{c}}, M.{\"U}., Yuce, F.: Comparison of five
  convolutional neural networks for predicting osteoporosis based on mandibular
  cortical index on panoramic radiographs. Dentomaxillofacial Radiology
  \textbf{51}(6),  20220108 (2022)

\bibitem{wang2022freematch}
Wang, Y., Chen, H., Heng, Q., Hou, W., Savvides, M., Shinozaki, T., Raj, B.,
  Wu, Z., Wang, J.: Freematch: Self-adaptive thresholding for semi-supervised
  learning. arXiv preprint arXiv:2205.07246  (2022)

\bibitem{wani2022osteoporosis}
Wani, I.M., Arora, S.: Osteoporosis diagnosis in knee x-rays by transfer
  learning based on convolution neural network. Multimedia Tools and
  Applications pp. 1--25 (2022)

\bibitem{zagoruyko2016wide}
Zagoruyko, S., Komodakis, N.: Wide residual networks. In: British Machine
  Vision Conference 2016. British Machine Vision Association (2016)

\end{thebibliography}
\end{document}